# Moiré Synergy: An Emerging Game Changer by Moiré of Moiré

Xiangbin Cai[1], Weibo Gao[1,2,3]*

[1] Division of Physics and Applied Physics, School of Physical and Mathematical Sciences, Nanyang Technological University, Singapore 637371, Singapore.

[2] The Photonics Institute and Centre for Disruptive Photonic Technologies, Nanyang Technological University, Singapore 637371, Singapore

[3] Centre for Quantum Technologies, National University of Singapore, Singapore.

* Correspondence to W. G. at wbgao@ntu.edu.sg

**Abstract**

Moiré superlattices of tunable wavelengths and the further developed moiré of moiré systems, by artificially assembling two-dimensional (2D) van der Waals (vdW) materials as designed, have brought up a versatile toolbox to explore fascinating condensed mater physics and their stimulating physicochemical functionalities. In this Perspective, we briefly review the recent progress in the emerging field of moiré synergy, highlighting the synergetic effects arising in distinct dual moiré heterostructures of graphene and transition metal dichalcogenides (TMDCs). A spectrum of moiré of moiré configurations, the advanced characterization and the exploitation efforts on the moiré-moiré interactions will be discussed. Finally, we look out for urgent challenges to be conquered in the community and some potential research directions in the near future.

Although the experimental investigation[1] and the theoretical prediction[2] of two-dimensional (2D) moiré properties followed closely the discovery of monolayer graphene[3] in 2004, the community enthusiasm towards the rich low-energy physics embedded in the moiré materials was not fully ignited, until Cao's demonstration of unconventional superconductivity[4] and correlated insulator[5] behaviors in the magic-angle twisted bilayer graphene (TBG) in 2018. Such moiré patterns in 2D heterostructures naturally form when two flakes of van der Waals (vdW) materials, including graphene and transition metal dichalcogenides (TMDCs), are stacked on top of each other with either a relative twist angle (for both homo- and hetero-bilayers) or a lattice mismatch (for hetero-bilayers). The prominent interlayer interactions[6-7] make the constituent atomic orbitals interfere constructively to manifest a superlattice of much larger wavelengths than the component lattice parameters, so that the energy band edges significantly flatten in the reciprocal space, nourishing a plethora of exotic quantum phenomena, such as superconductivity[8], correlated insulators[9], orbital magnetism[10] and unusual ferroelectricity[11].

Since the twist angle and the degree of lattice mismatch can be precisely and continuously tuned by the "tear-and-stack" micro-manipulation techniques and the synthetic adjustment of chemical compositions, respectively, scientists are now able to control the periodicity and the potential depth profile of each moiré superlattice at will.[12-13] Therefore, these years have witnessed that the moiré materials become the fertile playground for strong correlation physics. For example, the Hubbard model is well simulated by triangular moiré superlattices with controlled kinetic and potential terms.[14] And the single moiré properties have been accessed through a wide range of probing techniques, including electrical transport[4-5], thermodynamic[15] and optical measurements[16-17]. However, the simplicity and limited number of controllable degrees of freedom in a single moiré prevent the quantum simulation of other important models in condensed matter physics, such as the Bose-Hubbard model for strongly interacting bosons[18-19] and the Kondo lattice model that may require coupling two moiré bands of varied bandwidths[20].

Most recently, some pioneering works put their eyes on the landscape involving dual moirés as the functioning entity, where enhanced optoelectronic device characteristics[21-24] and unseen quantum phases[25-27] are revealed. Here in this Perspective, we briefly introduce the fabrication strategies of dual moiré systems and highlight the recent progress in the synergetic effects originating from the moiré of moiré heterostructures by graphene and popular TMDCs. We also give some heed to the advanced characterization techniques, which can help us better understand the structure-property relationship behind moiré-moiré interactions. We then finish with an outlook for the underlying challenges and opportunities in this infancy field.

**Intra- and Inter-moiré Configurations**

In general, regardless of the space symmetry of component sheets, an interfacial 2D moiré pattern appears when two distinct lattices of long-range orders overlap on top of each other, which is the rigid geometric interference of two lattices by their rotational and lateral phase differences at the interface. Regarding the most investigated vdW materials, i.e. graphene and TMDCs, both of them possess the honeycomb layered structure. The single moiré wavelength thus simply depends on the relative crystalline orientation (twist angle, $\theta$) and the variation in lattice parameters (lattice mismatch, $\Delta a$) between layers. As the twist angle and the lattice mismatch can be precisely and continuously manipulated by mechanical transfer and chemical synthesis, respectively, there are three design routes towards the targeted moiré of moiré heterostructures as illustrated in Figure 1: (1) physically stacking the same 2D vdW material flakes with consecutive (chiral) or alternative (symmetric) twisting schemes; (2) chemically changing the compositions of constituent layers, such as TMDC alloy with a full doping phase diagram[28], for desired interlayer lattice mismatches; (3) combining physical and chemical techniques for complicated heterostructures. It should be noted that the moiré of moiré system can actually be treated as the rigid geometric interference of two moirés, which indicates that the inter-moiré configuration, such as slight sliding or twisting between coupled moirés, plays a critical role for the final apparent superlattice. Therefore, similar to the single moiré scenario[29-31], the moiré periodicity of twisted homo-trilayers is more sensitive to the inter-moiré twisting while the final moiré lengths of hetero-trilayers are more robust against inter-moiré misalignment.

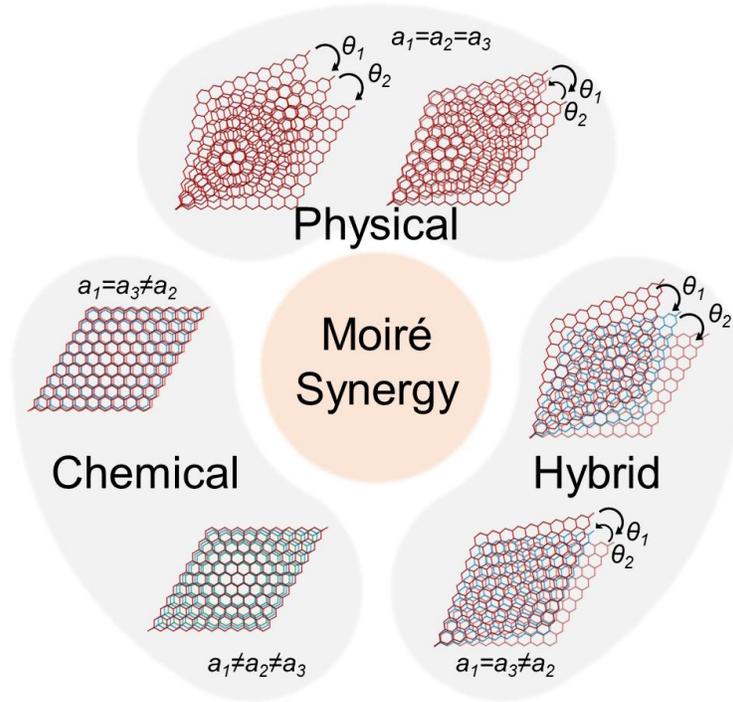

**Figure 1: Design routes for moiré of moiré systems.** Fabrication of dual moiré heterostructures can be achieved through **(top panel)** physically stacking the same 2D vdW material flakes with consecutive or alternative twisting schemes, **(left panel)** chemically changing the compositions of constituent layers for desired interlayer lattice mismatches or **(right panel)** the hybridization of these two techniques.

**Moiré-Moiré Interactions**

First of all, we need adequate knowledge about what we have fabricated, such as the crystalline symmetry, the interlayer coupling and the electronic structure of designed moiré-of-moiré systems, before we try to understand the observed unconventional phenomena stemming from moiré-moiré interactions. As one of the examples that apply advanced characterization techniques to moiré-of-moiré materials, in Figure 2 (a), Li *et al.* used conductive atomic force microscopy (C-AFM) to study consecutively twisted graphene trilayers and discovered unexpected symmetry breaking beyond the rigid model at neighboring domains of the large moiré.[32] Unlike TBG, the moiré-of-moiré graphene trilayers manifest anomalous conductivity at the AA stacking sites. Besides, the scanning probe microscopy based on AFM can also be employed to detect the piezo response in ferroelectric moirés.[33-34] Even though the angle-resolved photo-emission spectroscopy (ARPES) has a much worse spatial resolution than the AFM techniques do, usually few micrometers, it is an indispensable and powerful tool for visualizing the interlayer interactions of vdW materials. In Figure 2 (b), Xie *et al.* reported an imprinted large moiré potential in the graphene/$WS_2$/$WSe_2$ heterostructures, suggesting strong energy band hybridization between layers.[35] Combining atom-resolved scanning tunneling microscopy (STM), ARPES technique is promising to unravel the flat bands and electronic reconstruction in moiré-of-moiré materials.[36-37] In addition to these static probing techniques, by driving the band crossings between Dirac cone Landau levels and energy gaps in certain systems as demonstrated in Figure 2 (c), Shen *et al.* demonstrated the interesting Dirac spectroscopy in twisted graphene trilayers to extract the correlated gap size at a range of electric field strength.[38] The technique provides an exciting blueprint for investigating the complex phase transitions in moiré-of-moiré systems hosting both steep and flat bands.

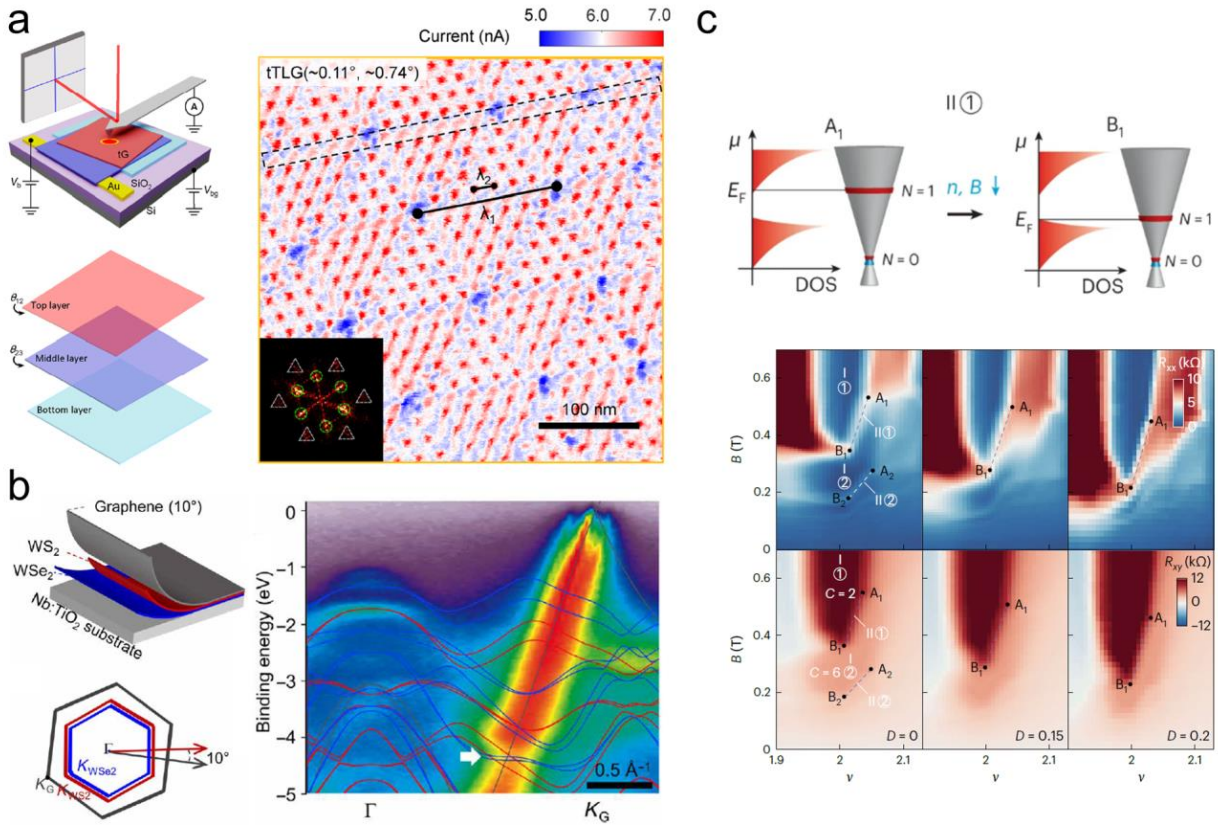

**Figure 2: Advanced characterization techniques on synergetic moiré-of-moiré materials. (a)** Conductive atomic force microscopy (C-AFM) study on consecutively twisted graphene trilayers. Reprinted with permission from ref. [32]. Copyright 2022 American Chemical Society. **(b)** Angle-resolved photo-emission spectroscopy (ARPES) study on aligned graphene/$WS_2$/$WSe_2$ trilayers. Reprinted with permission under a Creative Commons (CC BY) License from ref. [35]. Copyright 2022 American Association for the Advancement of Science. **(c)** Dirac spectroscopy study on the magic-angle twisted trilayer graphene. Reprinted with permission under a Creative Commons (CC BY) License from ref. [38]. Copyright 2022 Springer Nature.

The above structural knowledge of moiré-moiré interactions enables us to design specific multi-moiré heterostructures for their synergetic effects and understand the origins through a structure-property-correlated perspective of view. On one hand, the parallel multiple channels in moiré-of-moiré devices can stabilize and even promote the electronic and optoelectronic properties of single moirés. In Figure 3 (a), Zhang *et al.* found that with the number of alternatively twisted graphene layers increasing from three to five, superconductivity dome expands over a gradually enhanced filling-factor range, finally goes beyond the filling of four electrons per moiré super cell.[24] The critical electrical field strength to "turn off" such superconducting state also becomes reachable, which means the superconductivity is more controllable in multi-moiré systems. Interestingly, Xin *et al.* presented the prolonged exciton lifetime in an $WSe_2/WS_2/MoS_2$ heterostructure due to the lower exciton binding energy than that in the usual $WSe_2/MoS_2$ heterobilayer, as shown in Figure 3 (b).[23] In another report, as adapted in Figure 3 (c), the addition of a second moiré significantly improve the photoluminescence quantum yield in $WSe_2/MoSe_2/WSe_2$ heterostructures by a larger electron-hole wavefunction overlap in the dual moiré system, which suggests a shorter lifetime.[22] These two conclusions may look controversial but are actually consistent with their respective inter-moiré configurations. For $WSe_2/WS_2/MoS_2$ hetero-trilayers, dual moirés were constructed with the inversion symmetry broken, allowing for more gradual step-by-step transfer of charges, while the $WSe_2/MoSe_2/WSe_2$ hetero-trilayers consist of a symmetric homo-moiré configuration, where the electron layer is encapsulated by hole layers on both sides.

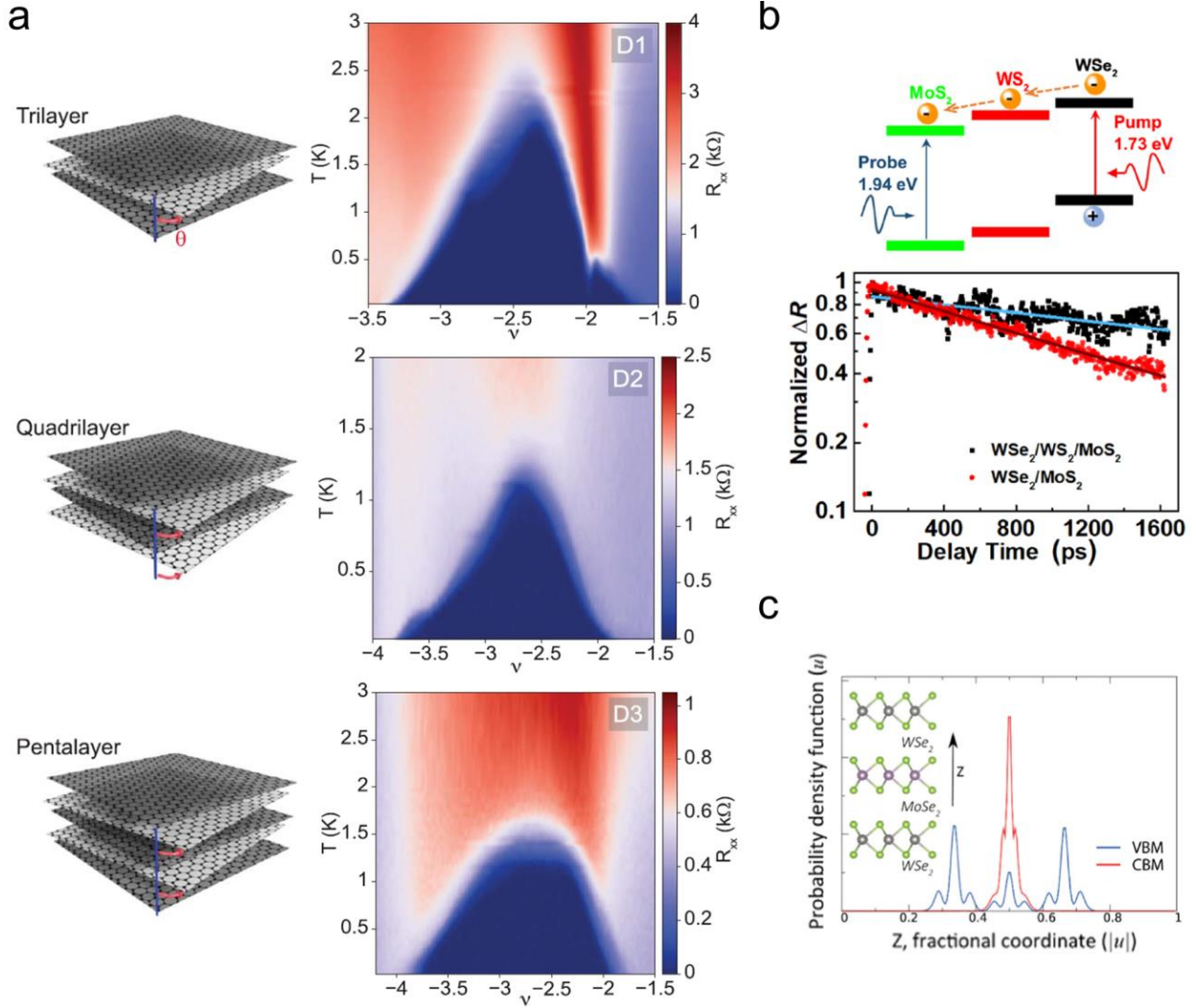

**Figure 3: Optoelectronic enhancement in synergetic moiré-of-moiré materials. (a)** Promotion of superconductivity in alternatively magic-angle twisted tri-, quadri- and pentalayer graphene. Reprinted with permission from ref. [24]. Copyright 2022 American Association for the Advancement of Science. **(b)** Lifetime improvement in the $WSe_2/WS_2/MoS_2$ heterostructures. Reprinted with permission from ref. [23]. Copyright 2022 AIP Publishing. **(c)** First-principles pseudopotential calculations of the increased electron-hole overlap in $WSe_2/MoSe_2/WSe_2$ trilayers. Reprinted with permission under a Creative Commons (CC BY) License from ref. [22]. Copyright 2018 Springer Nature.

On the other hand, the interference of wave functions in synergetic moiré-of-moiré materials can nurture new electronic structures that host quantum matter phases beyond what single moirés can do. In Figure 4 (a), Zhang *et al.* fabricated the consecutively twisted trilayer graphene of a higher-order moiré-of-moiré superlattice and observed correlated insulating and superconducting states at an extremely low carrier density (~$10^{10}$ cm$^{-2}$), which is different from the TBG case.[39] It is worth noting that the consecutively twisted trilayer graphene behaves quite distinctly from the alternatively twisted one[40], maybe owing to its larger moiré-of-moiré superlattice and thus flatter band edge in the Brillouin zone, which is another good example for the crucial role played by the inter-moiré configuration. In contrast to gapless graphene, moirés of semiconducting TMDCs can trap excitons, which are bosons of bound electron-hole pairs. With Coulomb-coupled moiré lattices in the angle-aligned $WS_2$/bilayer $WSe_2$/$WS_2$ multilayers, Zeng *et al.* unraveled intriguing exciton density waves in a Bose-Fermi mixture of excitons and holes as adapted in Figure 4 (b)[27], exhibiting excitonic insulating states at fractional electron filling factors that disappear in either single moiré. Similarly by confining excitons and charges in opposite moiré lattices, as shown in the calculated potential maps in Figure 4 (c), Li *et al.* demonstrated a charge-order sensing scheme utilizing the 1°/4° twisted $WSe_2$/$MoSe_2$/$WSe_2$ heterostructure[25]. A spectrum of correlated states at fractional fillings were detected with some of them manifesting spin characteristics.

These strong correlation effects in moiré-of-moiré systems share basic temperature and magnetic responses as those in single moirés, yet containing complexity and more controllable degrees of freedom. We have to admit that the field of synergetic moirés is still lacking in a general theoretic model accounting for the moiré-moiré interactions to guide the rational heterostructure design and the quantum phase discovery at this early stage, but some plausible specialized efforts have been noticed.[18-19, 41]

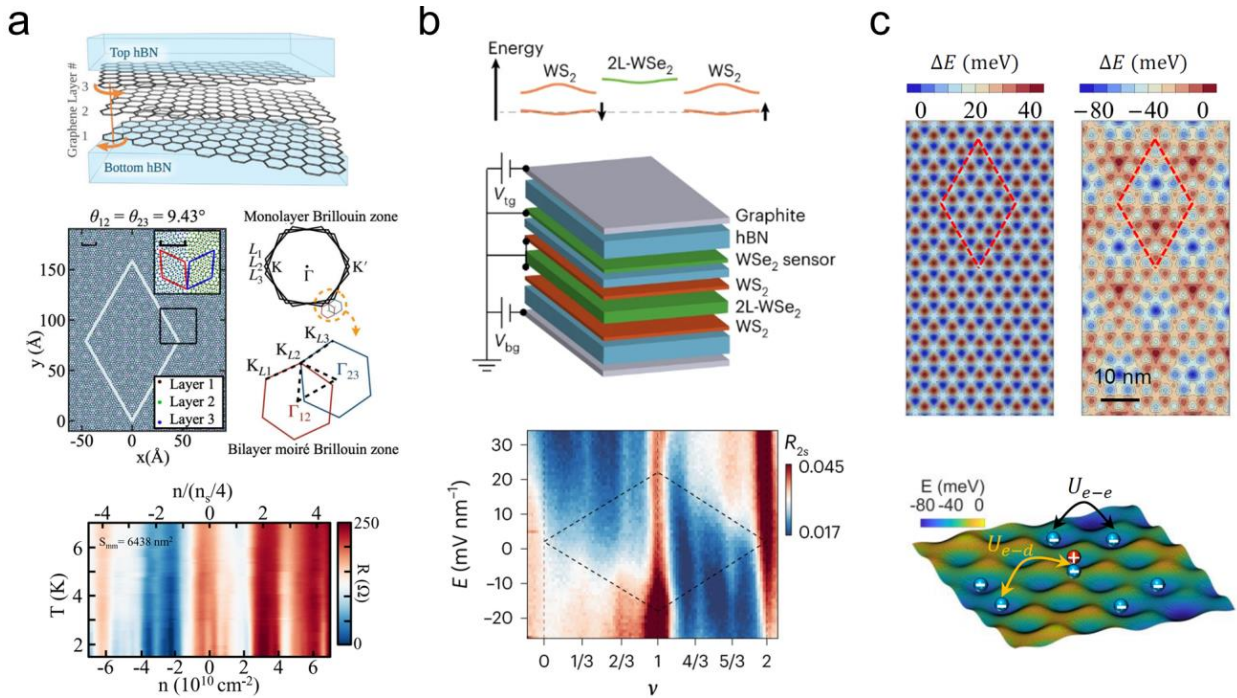

**Figure 4: Discovery of quantum phases in synergetic moiré-of-moiré materials. (a)** Correlated insulating states and superconductivity signature at an extremely low carrier density found in consecutively twisted trilayer graphene. Reprinted with permission from ref. [39]. Copyright 2021 American Physical Society. **(b)** Correlated insulating states at fractional filling factors originating from the exciton density waves in the angle-aligned $WS_2$/bilayer $WSe_2$/$WS_2$ multilayer. Reprinted with permission from ref. [27]. Copyright 2023 Springer Nature. **(c)** Calculated bottom exciton (left), middle electron potential (right) and the charge-order sensing scheme (bottom) in a 1°/4° twisted

WSe$_2$/MoSe$_2$/WSe$_2$ heterostructure. Reprinted with permission under a Creative Commons (CC BY) License from ref. [25]. Copyright 2021 the authors.

**Challenges and Opportunities**

Because the moiré of moiré system inevitably holds more interfaces and thus more degrees of freedom for careful consideration than those in single moiré, the challenges in device fabrication, structural characterization and optoelectronic property modeling are multiplied dramatically.

Interface cleanness and homogeneity: As Nobel laureate Herbert Kroemer put it, "the interface is the device", stressing the significance of interface engineering in constructing (opto-)electronic devices of thin semiconductor films. Such statement is even more the case for 2D material devices, in which the interface component takes up an ever larger proportion of the functioning entity in 2D heterostructures. Therefore, the interface cleanness and homogeneity remain one of the key aspects for fabricating a high-performance 2D device, especially the devices made up of moiré superlattices.[42-43] However, contaminants from sample exfoliation, transfer and stacking processes are easily trapped between layers, aggregating into blisters in each interface as scattering centers and disturbing the interlayer alignment for long-range atomic order. There have been lots of efforts in mechanically cleaning the assembled vdW interface, such as hot pick-up[44], high-temperature laminating[45] and AFM tip cleaning[46] techniques. But these methods are demanding in intricate workmanship and not scalable for industrial purposes. Ideally, the mass production of moiré materials by chemical vapor deposition (CVD) synthesis is cost-effective and promising for high-quality heterostructures of atomic-level clean and uniform interfaces.[47-48] Unfortunately, the traditional CVD requires a high temperature of 550-1000 °C for the reliable growth of crystalline 2D materials[49], which hinders the sequential growth of component layers of moiré heterostructures due to the thermal-induced degradation. Therefore, the low-temperature synthetic strategy for high-crystallinity large-scale multi-moiré heterostructures can be one of prospective directions.

Electrical contact: Any progress in either the fundamental research or device applications of moiré materials crucially relies on the advances in sample fabrication techniques, including the interfaces inside heterostructures as discussed above and the interfaces between channel materials and external circuitry. A robust Ohmic electrical contact is desired for the efficient charge carrier injection in metal-semiconductor junctions. In practice, however, the Fermi level pinning (FLP) effects naturally arise when there are chargeable states (either metal-induced gap states or surface states) in the semiconductor energy gap. This results in the semiconductor's band bending and the notorious Schottky barrier, in addition to the tunnel barrier from vdW gap.[50] Although substantial efforts have attempted to improve the quality of metal-TMDC electrical contacts, such as through depositing orbital-hybridized semi-metals as electrodes[51-52], inserting thin tunnel barriers[53], and fabricating vdW-contacted metals[54-55], just to name a few, a universal electrical contact strategy of high reproducibility and scalability, which are actually weak in most of similar works, for different moiré materials still remains highly desirable.

Theoretic modeling with interface relaxation included: The weak vdW interactions in 2D materials not only provide unlimited possibilities in engineering artificial layered solids, but also introduce trouble to the accurate theoretic modeling of vdW heterostructures. Currently the simplified moiré band model was built upon the rigid geometry stacking of vdW layers without any lattice relaxation or interface reconstruction[56-57], which yet even exists in multilayers[58]. For unreconstructed interfaces, which assumption greatly reduces the structure complexity and the computation load, the local stacking within a moiré super unit cell continuously shifts like a rigid solid, in regardless of the intralayer distortion and built-up stress in the real 2D sheets, which may significantly alter the band structure and apparent properties. In the heterostructure composed of flexible vdW layers of few-atomic-layer thickness, the interlayer adhesion varies with the local atomic registry and finally achieves the cost-effective equilibrium against the elastic intralayer deformation.[30] Such interface reconstruction and strain modulation in moiré of moiré systems should be more complicated and substantially modify the electronic structure, though the corresponding general theories are still missing.

Despite of challenges and difficulty faced by scientists, numerous opportunities also exist in this rising field. We propose a few promising directions in exploiting the moiré synergy below.

Multiferroics: For decades, ferroelectricity and ferromagnetism had been considered as collective phenomena in bulky materials, until the recent boom of vdW materials and their heterostructures, especially the unconventional ferroelectricity[33-34, 59] and ferromagnetism[60] in moiré superlattices of non-ferroic layers. It is exciting if we could couple ferroelectric and ferromagnetic moirés as multiferroic devices through the moiré of moiré schemes. Recently there have also been reports on the phase-controlled synthesis of ferroic vdW materials[61-62] and the fabrication of moiré-of-moiré heterostructure by ferroelectric and ferromagnetic atomic layers can also generate artificial multiferroic devices. Through the realization and study of multiferroic double-moiré devices, we should be able to get deeper understanding on the fundamental mechanisms of ferroic moirés and provide mixed-ferroic possibilities for future information storage.

*In-situ* inter-moiré tunability and localized probes: In moiré of moiré heterostructures, the inter-moiré configurations, either sliding or twisting, markedly affect the apparent electronic properties. The moiré-moiré interactions can be precisely and continuously tuned by changing the inter-moiré stacking order. There have been some proof-of-concept demonstrations on the *in-situ* dynamic rotation of single moiré devices[63-64], which techniques should be transferrable to the multi-moiré system to achieve continuous tuning of the inter-moiré interaction strength. This will allow us to study the quantum phase transition with unprecedented control. At the same time, with a highly localized probe that can access the charge, spin and collective excitations of moiré systems, such as sub-angstrom electron probe[65], nanoscale scanning superconducting quantum interference device (nano-SQUID)[42] and high-resolution electron energy-loss spectroscopy (HR-EELS)[66], we can investigate the underlying quantum phase behaviors in synergetic moiré-of-moiré materials at an unparallel level.

Heterodimensional superlattice: Integrating materials of different dimensionalities usually forms unique superlattices of unconventional properties[67] and provide a viable scheme for device design. Scientists have found that periodically patterned dielectric substrates can effectively modulate the electronic properties of supported 2D materials beyond the constraints of atomic crystallinity.[68-69] In contrast to 2D crystals of fixed lattice parameters, both periodicities of 2D moirés and patterned substrates can be adjusted, so that the 2D moiré on the patterned dielectric substrate can form a unique moiré of moiré system. When the two periodicities are tuned, we expect to uncover novel electronic properties residing in this heterodimensional superlattice.

**Conclusion**

To sum up, with the advent of vdW heterostructures, moiré superlattices and moiré of moiré systems offer an unprecedented playground to play with fascinating condensed matter physics and integrate diverse functionalities at the low dimensionality, especially the quantum simulations of strong correlation models. While challenges to be conquered, there have been already fruitful and motivating achievements so far, and we see numerous opportunities in this emerging field of moiré synergy.

**Additional Information:** The authors declare no competing financial interests.

**Acknowledgement:** This work is supported by the Competitive Research Program of Singapore National Research Foundation (CRP Awards No. NRF-CRP22-2019-0004, No. NRF-CRP23-2019-0002) and the Singapore Ministry of Education (MOE2016-T3-1-006 (S)).